\documentclass[%
twocolumn,
superscriptaddress,
 amsmath,amssymb,
 aps,
 prl,
 showpacs,
floatfix,
]{revtex4-1}

\usepackage[utf8]{inputenc}
\usepackage{xcolor}
\usepackage{amsmath}
\usepackage{amssymb}
\usepackage{graphicx}
\usepackage{mathtools, cuted}
\usepackage{titlesec}
\usepackage{parskip}
\usepackage{flushend}
\usepackage{float}

\newcommand{\Yb}{$\mathrm{^{171}Yb}$}
\newcommand{\spinup}{|{\uparrow}\rangle}
\newcommand{\spindown}{|{\downarrow}\rangle}

\newcommand{\un}[1]{\mathrm{\:#1}} 
\newcommand{\term}[3]{^{#1} \hspace{-0.1 em} #2_{#3}} 

\newcommand{\ket}[1]{\left|  #1 \right\rangle}

\newcommand{\aver}[1]{\ensuremath{\langle {#1} \rangle}}
\newcommand{\Nup}[0]{\ensuremath{N_\uparrow}}
\newcommand{\Ndown}[0]{\ensuremath{N_\downarrow}}
\newcommand{\var}[1]{\ensuremath{\left( \Delta #1 \right)^2}}

\definecolor{plotgreen}{RGB}{0,150,0}

\renewcommand\thefigure{\arabic{figure}}
\renewcommand\theequation{\arabic{equation}}
\usepackage{graphicx}
\usepackage{dcolumn}
\usepackage{bm}
\usepackage{hyperref}
\usepackage[mathlines]{lineno}

\usepackage{xcolor}
\usepackage{amssymb}
\usepackage{verbatim} 
\usepackage{comment}
\usepackage{gensymb} 
\usepackage{soul} 
\usepackage{subfiles}
\usepackage{amsmath}
\usepackage{mathtools, cuted}
\usepackage{xr}
\usepackage{graphicx}

\definecolor{plotgreen}{RGB}{0,150,0}

\begin{document}

\title{Near-Unitary Spin Squeezing in $^{171}$Yb}

\author{Boris Braverman}
\email{bbraverm@uottawa.ca}
\altaffiliation{Current address: Department of Physics and Max Planck Centre for Extreme and Quantum Photonics, University of Ottawa, 25 Templeton Street, Ottawa, Ontario K1N 6N5, Canada}
\affiliation{Department of Physics, MIT-Harvard Center for Ultracold Atoms and Research Laboratory of Electronics, Massachusetts Institute of Technology, Cambridge, Massachusetts 02139, USA}

\author{$^\ddagger$ Akio Kawasaki}
\email{akiok@stanford.edu}
\altaffiliation{Current address: W. W. Hansen Experimental Physics Laboratory and Department of Physics, Stanford University, Stanford, California 94305, USA}
\affiliation{Department of Physics, MIT-Harvard Center for Ultracold Atoms and Research Laboratory of Electronics, Massachusetts Institute of Technology, Cambridge, Massachusetts 02139, USA}

\author{$^\ddagger$ Edwin Pedrozo-Pe\~{n}afiel}%
\thanks{These authors contributed equally}
\affiliation{Department of Physics, MIT-Harvard Center for Ultracold Atoms and Research Laboratory of Electronics, Massachusetts Institute of Technology, Cambridge, Massachusetts 02139, USA}

\author{Simone Colombo}%
\affiliation{Department of Physics, MIT-Harvard Center for Ultracold Atoms and Research Laboratory of Electronics, Massachusetts Institute of Technology, Cambridge, Massachusetts 02139, USA}

\author{Chi Shu}%
\affiliation{Department of Physics, MIT-Harvard Center for Ultracold Atoms and Research Laboratory of Electronics, Massachusetts Institute of Technology, Cambridge, Massachusetts 02139, USA}
\affiliation{Department of Physics, Harvard University, Cambridge, Massachusetts 02138, USA}

\author{Zeyang Li}%
\affiliation{Department of Physics, MIT-Harvard Center for Ultracold Atoms and Research Laboratory of Electronics, Massachusetts Institute of Technology, Cambridge, Massachusetts 02139, USA}

\author{Enrique Mendez}%
\affiliation{Department of Physics, MIT-Harvard Center for Ultracold Atoms and Research Laboratory of Electronics, Massachusetts Institute of Technology, Cambridge, Massachusetts 02139, USA}

\author{Megan Yamoah}%
\affiliation{Department of Physics, MIT-Harvard Center for Ultracold Atoms and Research Laboratory of Electronics, Massachusetts Institute of Technology, Cambridge, Massachusetts 02139, USA}

\author{Leonardo Salvi}%
\affiliation{Department of Physics, MIT-Harvard Center for Ultracold Atoms and Research Laboratory of Electronics, Massachusetts Institute of Technology, Cambridge, Massachusetts 02139, USA}
\affiliation{Dipartimento di Fisica e Astronomia and LENS - Universit\`{a} di Firenze, INFN - Sezione di Firenze, Via Sansone 1, 50019 Sesto Fiorentino, Italy}

\author{Daisuke Akamatsu}%
\affiliation{Department of Physics, MIT-Harvard Center for Ultracold Atoms and Research Laboratory of Electronics, Massachusetts Institute of Technology, Cambridge, Massachusetts 02139, USA}
\affiliation{National Metrology Institute of Japan (NMIJ), National Institute of Advanced Industrial Science and Technology (AIST), 1-1-1 Umezono, Tsukuba, Ibaraki 305-8563, Japan}

\author{Yanhong Xiao}%
\affiliation{Department of Physics, MIT-Harvard Center for Ultracold Atoms and Research Laboratory of Electronics, Massachusetts Institute of Technology, Cambridge, Massachusetts 02139, USA}
\affiliation{Department of Physics, State Key Laboratory of Surface Physics and Key Laboratory of Micro and Nano Photonic Structures (Ministry of Education), Fudan University, Shanghai 200433, China}

\author{Vladan Vuleti\'{c} }
\email{vuletic@mit.edu}
\affiliation{Department of Physics, MIT-Harvard Center for Ultracold Atoms and Research Laboratory of Electronics, Massachusetts Institute of Technology, Cambridge, Massachusetts 02139, USA}

\date{\today}

\begin{abstract}
Spin squeezing can improve atomic precision measurements beyond the standard quantum limit (SQL), and unitary spin squeezing is essential for improving atomic clocks. We report substantial and nearly unitary spin squeezing in $^{171}$Yb, an optical lattice clock atom. The collective nuclear spin of $\sim 10^3$ atoms is squeezed by cavity feedback, using light detuned from the system's resonances to attain unitarity. The observed precision gain over the SQL is limited by state readout to 6.5(4) dB, while the generated states offer a gain of 12.9(6) dB, limited by the curvature of the Bloch sphere. Using a squeezed state within 30\% of unitarity, we demonstrate an interferometer that improves the averaging time over the SQL by a factor of 3.7(2). In the future, the squeezing can be simply transferred onto the optical clock transition of $^{171}$Yb.

\end{abstract}

\maketitle

Optical lattice clocks (OLCs) employ ensembles of cold trapped atoms to reach unprecedented fractional accuracy at the level of $10^{-18}$ \cite{Ludlow2015,Ushijima2015,Campbell2017,McGrew2018,Marti2018}. Such clocks now operate near the standard quantum limit (SQL) set by quantum projection noise, where the precision of a sensor improves as $\sqrt{N}$ with the number of atoms $N$. Spin squeezed states (SSSs) \cite{Kitagawa1993,wineland1992a,Wineland1994,ma2011quantum,Takano2009,Appel2009,esteve2008,Riedel2010,Leroux2010,Schleier-Smith2010a,Hamley2012,Bohnet2016,Bao2018,Cox2016a,Hosten2016,Hosten2016a,Sewell2012}
are many-body entangled states that can overcome the SQL \cite{Wineland1994,Pezze2018}. They have simple Gaussian quasi-probability distributions with reduced (squeezed) and enhanced (antisqueezed) quantum noise, respectively, along two orthogonal directions of the collective atomic spin. While for fixed-bandwidth applications the precision depends on the squeezing alone, \citet{Andre2004} have shown that for optimized clocks the antisqueezed direction eventually leaks into the measurement, reducing the gain in precision. In practice,  the amount of antisqueezing typically far exceeds the squeezing, and this mechanism can dramatically reduce the precision gain to the point where, e.g., the state with the highest inferred squeezing of 20~dB (and an antisqueezing of 39~dB) \cite{Hosten2016} would improve the precision of a clock by a mere 2~dB \cite{Braverman2018}. Thus nearly unitary (area-preserving) squeezing is of high importance for future clock applications. Furthermore, of the most common OLC atoms, spin squeezing in Sr, Ca, Mg or Hg have not been demonstrated so far, and Yb has only been weakly squeezed by $\sim 2$~dB \cite{Takano2009}.

In this Letter, we demonstrate for the first time near-unitary optical spin squeezing, as well as the first substantial squeezing in an OLC atom. The observed metrological gain of up to $6.5(4)\un{dB}$ is limited by the state detection, while subtraction of the independently determined measurement noise implies that the generated SSSs offer $12.9(6)\un{dB}$ of metrological gain and $15.9(6)\un{dB}$ of spin noise suppression. Under conditions where the squeezing is unitary within 30\%, and nearly optimal for clock applications, we demonstrate an interferometer with a factor of 3.7(2) reduction in averaging time over the SQL. In the future, the demonstrated squeezing between the two nuclear sublevels $\ket{m=\pm\frac{1}{2}}$ of the electronic ground state $^1S_0$ of \Yb~ can be directly used in the OLC by transferring the population of one of the two sublevels into the $^3P_0$ excited clock state with an optical $\pi$ pulse \cite{Lemke2009}.

\begin{figure}[hbtp]
\setlength{\unitlength}{1\textwidth}
\includegraphics[width = \columnwidth]{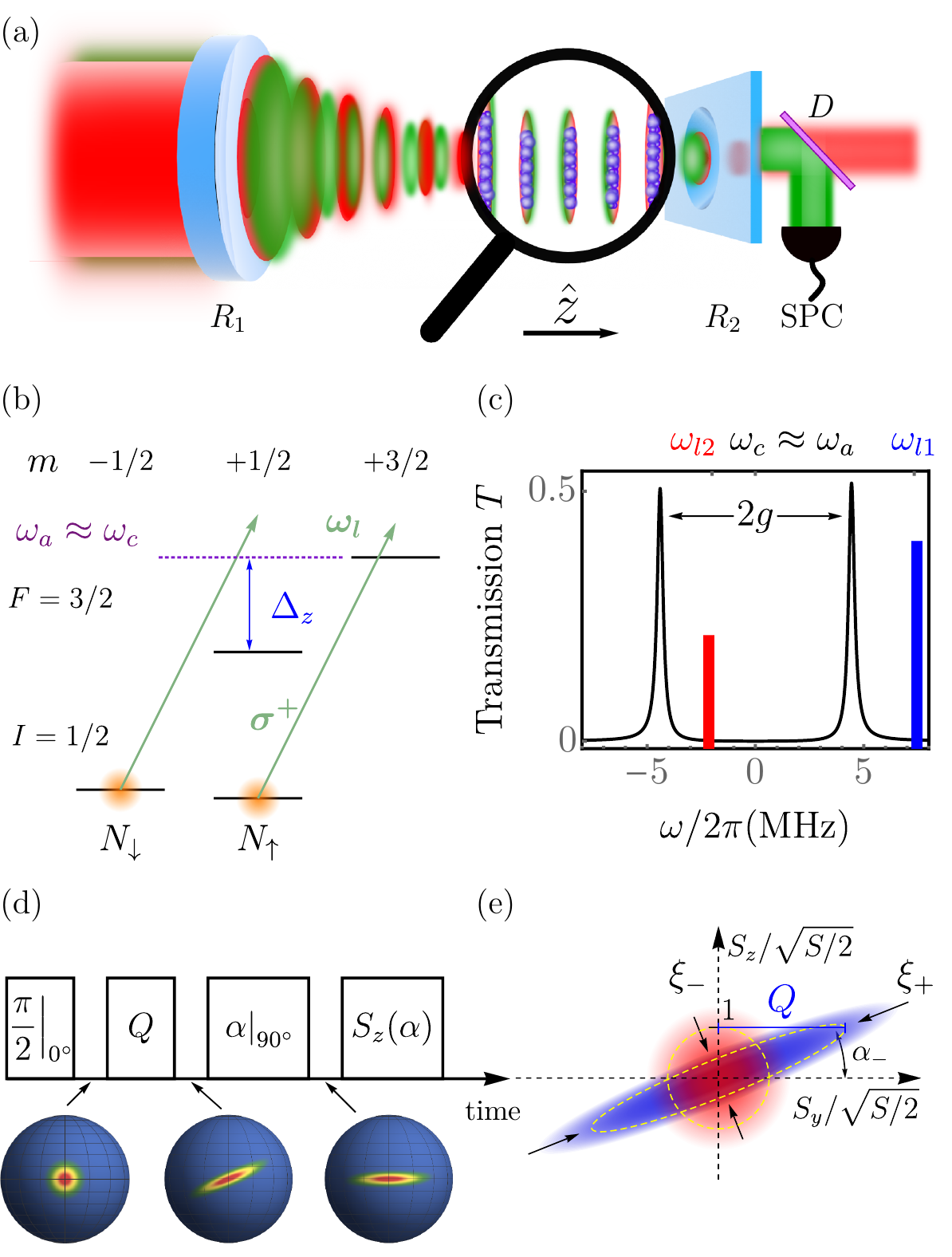}
\caption{(a) Experimental setup. A one-dimensional optical lattice at $\lambda_{t}{=}759 \un{nm}$ (red) traps the atoms. Light at $\lambda{=}556 \un{nm}$ light (green), whose transmission is detected using a dichroic mirror ($D$) and a single-photon counter (SPC), is used for squeezing and probing. 
(b) Relevant energy levels of $^{171}$Yb with ground state $\ket{^1S_0, I=\frac{1}{2}}$ and excited state $\ket{^3P_1, F=\frac{3}{2}}$. The Zeeman splitting in $\ket{^3P_1, F=\frac{3}{2}}$ is $\Delta_{z}/(2\pi){=}18.5$~MHz for a magnetic field $B_z{=}13.6\un{G}$ along the cavity axis. (c) Cavity transmission spectrum showing vacuum Rabi splitting for $N\eta{=}1800$, as well as the two squeezing light pulses $\omega_{l1}$ and $\omega_{l2}$. (d) Simplified representation of the squeezing and measurement sequence. (e) Quasiprobability distributions for a CSS (red) and SSS (blue).}
\label{fig:Scheme}
\end{figure}

Optical spin squeezing methods rely on the collective interaction of the atomic ensemble with a light field, where for superior performance the atom-light interaction is enhanced by a cavity \cite{Schleier-Smith2010a,Cox2016a,Hosten2016}. One method that does not require detection of the light, which in practice is always imperfect, is cavity feedback squeezing \cite{Leroux2010,Schleier-Smith2010,Hosten2016a}: The spin quantum noise tunes the cavity frequency, such that the amount of light circulating inside the cavity depends on the $S_z$ component of the collective atomic spin. The light then acts back onto another component $S_y$ of the atomic spin through the light shift, creating $S_y$-$S_z$ quantum correlations and atomic entanglement in the process. In cavity squeezing, any information contained in the light field results in non-unitary evolution of the atomic system \cite{Leroux2012}. Recently, \citet{Zhang2015} pointed out that the process can be made more unitary by detuning the probe light far from cavity resonance. Although this decreases the squeezing strength per photon, it also hides the information about the atomic state in the photon shot noise, thereby enhancing the squeezing. In the present work, we are making use of this idea, but in a resonant regime of vacuum Rabi splitting, rather than dispersive cavity shift \cite{Zhang2015}, resulting in further improved squeezing, and more resilience to technical noise. In addition, we implement spin squeezing on an almost closed optical transition, which removes a squeezing limit due to Raman scattering between the spin states \cite{Saffman2009,Cox2016a}, and allows us for the first time to create SSSs that are limited by the curvature of the Bloch sphere for the collective atomic spin (see Fig. \href{fig:Scheme}).\\
Laser-cooled \Yb~atoms are prepared in a magic-wavelength optical-lattice trap inside an optical cavity. The atom-light interaction is characterized by an effective single-atom cooperativity $\eta{=}1.8(1)$ and collective cooperativity $N\eta{\approx}1800$, where $N{\approx}1000$ is the effective atom number (see Ref. \cite{Kawasaki2019} and Supplemental Material (SM)~\cite{SM} for details). The value of the effective cooperativity $\eta$ is confirmed in an independent measurement.

We perform squeezing between the nuclear sublevels $\spinup\equiv \ket{m_I{=}\frac{1}{2}}$ and $\spindown \equiv \ket{m_I{=}{-}\frac{1}{2}}$ of the electronic $\term{1}{S}{0}$ ground state of \Yb. The collective spin state can be represented on a Bloch sphere with radius $S{=}N{/}2$~\cite{Hu2015}. The cavity frequency is tuned to be nearly resonant with the $\spinup \rightarrow \ket{\term{3}{P}{1}, m_F=\frac{3}{2}}$ atomic transition. $\Nup$ atoms in the state $\spinup$ induce a vacuum Rabi splitting $2g{=}\sqrt{N_\uparrow \eta \kappa \Gamma}$ of the atom-cavity resonance (Fig.~\ref{fig:Scheme}(c)), where $\kappa$ and $\Gamma$ are the cavity and atomic linewidth, respectively. There is also a small dispersive effect from the $\Ndown$ atoms in the state  $\spindown$, suppressed by the Zeeman splitting in the excited $\term{3}{P}{1}$ state, with $\Delta_{z}{\gg}{\Gamma,\kappa}$ (see Fig.~\ref{fig:Scheme}(b)); this effect is included in our theoretical model (see SM \cite{SM} for details).

Since the cavity is primarily coupled to the population $\Nup$ of the state $\spinup$, $S_z$ is determined by detecting $\Nup$ via a measurement of the Rabi splitting $2g$, swapping the populations of $\spinup$ and $\spindown$ with a radiofrequency $\pi$ pulse, and remeasuring the Rabi splitting to give $\Ndown$. From $\Nup$ and $\Ndown$, we determine $S_z=(\Nup-\Ndown)/2$, and $S=(\Nup+\Ndown)/2$ using the two-transition atomic model and the separately measured cavity parameters (see SM~\cite{SM}).
The primary quantity of interest, denoted by $\sigma^2 \equiv 2\var{S_z}/S$, is the spin variance $\var{S_z}$ normalized to the noise of the coherent spin state (CSS) $\var{S_z}_{CSS}=S/2$. The SQL corresponds to $\sigma^2{=}1$.

The measured spin variance $\sigma^2$ is the sum of the variances of the atomic state $\sigma^2_{st}$ and the measurement resolution $\sigma_d^2$. To independently quantify the latter, we prepare a CSS on the equator, measure $S_z$ twice, and set $\sigma_d^2 \equiv \un{var}(S_{z1}-S_{z2})/2$. We achieve a detection variance $\sigma^2_d=-9.4(4)$~dB, i.e. a factor of $9$ below the SQL. The measurement quality is limited by a small residual Raman scattering that randomly transfers atoms between the states $\spinup$ and $\spindown$ \cite{Schleier-Smith2010,Hosten2016,Cox2016a,Saffman2009,Chen2014}, in combination with the collective cooperativity $N \eta$ and photon detection efficiency $\epsilon=15\%$  (see SM~\cite{SM}). 

The spin squeezing sequence is shown in Fig.~\ref{fig:Scheme}(d). First, we create a CSS along the $x$-axis by optically pumping all atoms into $\spinup$, and then applying a $\pi/2$-pulse. The squeezing is generated by pulses of light \cite{Leroux2010,Schleier-Smith2010a}, whose frequency $\omega_l$ is chosen to balance two competing effects: Increased detuning from the vacuum Rabi peaks makes the squeezing process more unitary with respect to the transmitted light, but also reduces the squeezing per photon and the interferometer contrast (see Fig.~\ref{fig:Allan}(b)). Furthermore, fluctuations in the trapped atom number result in fluctuations of the squeezing strength since the vacuum Rabi splitting depends on $\Nup$, rather than $S_z$. We cancel this effect by squeezing with bichromatic light inside and outside the Rabi peaks  (see Fig.~\ref{fig:Scheme}(c)), so that the combined squeezing is independent of total atom number. Besides the squeezing, the intracavity light also shifts the phase of the CSS, which we cancel by a spin echo sequence  with two bichromatic pulses (see SM~\cite{SM} for a detailed description). 

\begin{figure}[ht]
\setlength{\unitlength}{1\columnwidth}
 \begin{picture}(1,.66)
		\put(.0,0){\makebox{\includegraphics[width=0.99\columnwidth]{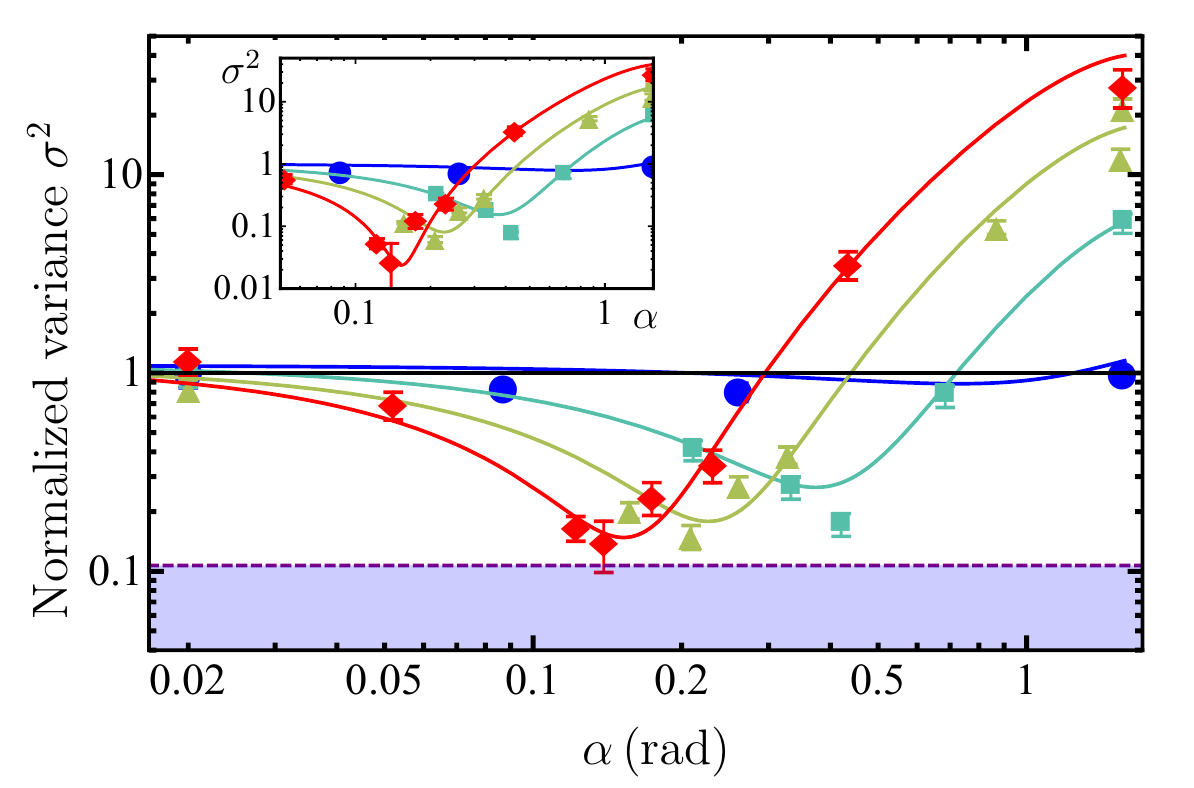}}}
    \end{picture}
\caption{Measured normalized spin noise $\sigma^2(\alpha)$, as a function of the state rotation angle $\alpha$, for shearing strengths $Q=0.3$ (blue circles), $Q=2.2$ (green squares), $Q=4.5$ (yellow triangles), and $Q=6.3$ (red diamonds). For visualization, the data measured at $\alpha{=}0$ are displayed at $\alpha{=}0.02$~rad. The solid lines are theoretical fits. States in the violet region below the dashed line at $\sigma^2=\sigma_d^2=0.11$ (detection limit) cannot be directly observed. Inset: $\sigma_{st}^2$ of SSS after subtracting measurement noise $\sigma_d^2$ for the same parameters.}
\label{fig:tomography}
\end{figure}

The generated SSS is reconstructed by rotating it by an angle $\alpha$ about its average spin vector and then detecting the spin projection $S_z(\alpha)$ along the $z$ axis (see Fig.~\ref{fig:Scheme}(d)). The measurement is repeated more than 100 times for each $\alpha$. The normalized spin variance $\sigma^2(\alpha){=}2 \var{S_z(\alpha)} /S$ along the direction $\alpha$ is displayed in Fig.~\ref{fig:tomography} for several different powers of the squeezing light. As a given SSS is rotated, the projected variance dips below the CSS noise until the rotation angle $\alpha$ reaches $\alpha_{-}$, where the short axis of the uncertainty ellipse lies along the $z$-axis. Beyond $\alpha_{-}$, the variance grows until the antisqueezing quadrature is oriented along $z$ for $\alpha= \alpha_{-}+\pi/2$.

To compare the data to a theoretical model, we first consider the polar angle of the spin vector, defined as $\tau_{\alpha} {\equiv} \sqrt{2S} \arcsin {\left(S_z(\alpha)/S \right)}$. Its variance $\var{\tau_{\alpha}}$, normalized to the variance $\var{\theta}_{CSS}=(2S)^{-1}$ of the CSS, is given by 
\begin{equation}
\frac{\var{\tau_{\alpha}}}{\var{\theta_{CSS}}} = 1-Q\sin{2\alpha}+\left(F+Q^2\right){\sin^2\alpha}.
    \label{eq:varSangular}
\end{equation}
Here, $Q$ is the dimensionless shearing strength (see Fig.~\ref{fig:Scheme}(e)), defined as the normalized light--induced phase shift ${\pm}\phi{/}\Delta\theta_{CSS}$ experienced by a spin displaced by one standard deviation of the CSS from the equator, $S_z=\pm \sqrt{S/2}$~\cite{Leroux2010}. The other dimensionless parameter $F$ quantifies the excess broadening (in variance units) compared to a pure SSS or CSS, which have $F=0$. (For an explicit expression for $Q$ and $F$ see SM~\cite{SM}.)

From Eq. \eqref{eq:varSangular} we find the minimum ($\xi_{-}^2$) and maximum ($\xi_{+}^2$) variances of the normalized spin angle $\tau_{\alpha}$, 
\begin{equation}
   \xi_{\pm}^2 = \frac{1}{2} \left( 2+F+Q^2 \pm \sqrt{4Q^2 + (F+Q^2)^2}\right), \label{eq:squeezing}
\end{equation}
obtained at angles $\alpha_{-}{=}\arctan[(\sqrt{4Q^2{+}(F{+}Q^2)^2}-(F{+}Q^2)){/}(2Q)]$ and $\alpha_{+}{=}\alpha_{-}{+}\pi/2$, respectively. The normalized uncertainty area of the SSS ellipse is given by $A{=}\xi_{+} \xi_{-}{=}\sqrt{1{+}F}$. The relation between $\var{\tau_{\alpha}}$ and the normalized spin variance $\sigma^2 (\alpha)$ of $S_z(\alpha)$ is given by $\sigma^2(\alpha){=} (S{/}2)[1-\textrm{exp}(-2\var{\tau_{\alpha}}{/}S)]$. For the CSS and the SSS quadrature $\xi_{-}^2$, the approximation  $\sigma^2(\alpha) {\approx} \var{\tau_{\alpha}}$ holds, while the antisqueezed quadrature $\xi_{+}^2$ is reduced by the curvature of the Bloch sphere.

\begin{figure}[hbtp]
\includegraphics[width=\columnwidth]{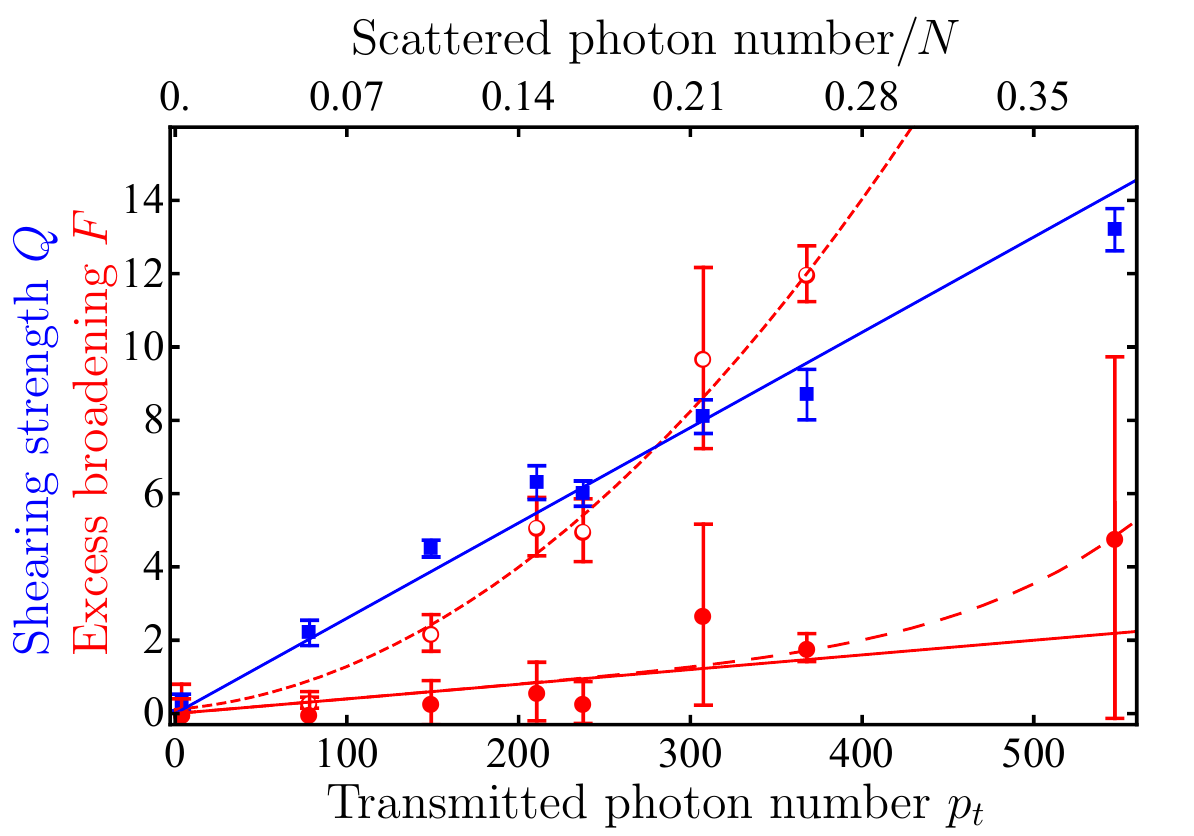}
\caption{Shearing strength $Q$ (filled blue squares) and excess broadening factor $F$ (red circles) plotted vs. the number $p_t$ of transmitted photons. The open red circles correspond to the directly measured data with theoretical model without free parameters (dotted red line), the solid circles are after subtraction of measurement noise $\sigma_d^2$ with parameter-free model without (solid red line) and with (dashed red line) Bloch-sphere-curvature-induced broadening. The theoretical predictions are given by Eqs. (S7) and (S8) of the SM~\cite{SM}.
}
\label{fig:QandF}
\end{figure}

\begin{figure}[ht]
\includegraphics[width = \columnwidth]{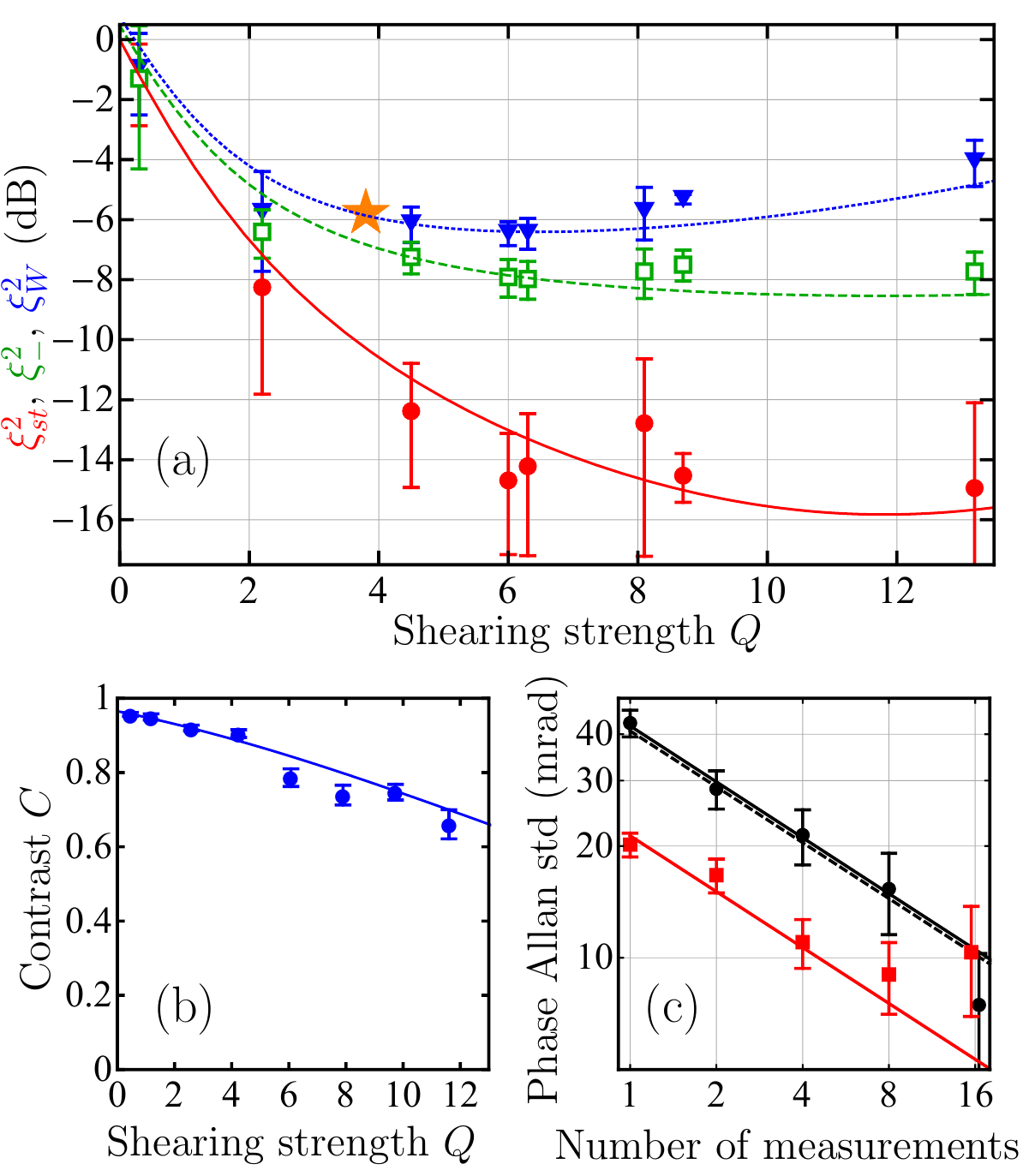}
\caption{(a) Wineland metrological gain $\xi_W^2$ (blue), measured spin noise reduction $\xi_{-}^2$ (green), and inferred SSS noise $\xi_{st}^2$ (red) as a function of shearing strength $Q$. $\xi_{-}^2$ is limited by the measurement resolution and $\xi_{st}^2$ by the curvature of the Bloch sphere. (b) Ramsey contrast as a function of $Q$ with initial contrast as only fitting parameter (solid line). (c) Allan deviation of a phase measurement for a CSS (black squares) with SQL (dashed line) and for a SSS with $Q=3.8~(p_t{=}130), F=0.8$ (red data). The red solid line is fit to the first three data points. The reduction in measurement time over the SQL is a factor of 3.7(2), represented by the orange star in (a).
}
\label{fig:Allan}
\end{figure}

The solid lines in Fig.~\ref{fig:tomography} are obtained by fitting the data to $\sigma^2(\alpha){+}\sigma^2_d$ with $Q$ and $F$ as the only fitting parameters, while $\sigma_d^2$ is the previously measured detection limit. We find good agreement between the model and the data, allowing us to extract both the shearing strength $Q$ and the excess broadening $F$. In Fig.~\ref{fig:QandF} we plot $Q$ and $F$ versus the number $p_t$ of transmitted photons during the optical squeezing. For negligible technical noise we expect both $Q$ and $F$ to be proportional to $p_t$ (see SM~\cite{SM}). The solid lines in Fig.~\ref{fig:QandF} represent the predicted linear behavior of $Q$ and $F$ obtained from an analytical model of the system without any free parameters (see SM~\cite{SM}); the dotted line includes the effect of finite measurement quality $\sigma_d^2{=}-9.4$~dB, that affects the measurement of the squeezed quadrature $\xi_{-}^2{<}1$, and hence $F$, but not $Q$. The model without any free parameters agrees remarkably well with the measured $Q$ and $F$, indicating the absence of major technical limitations other than the finite state detection quality $\sigma_d^2$.

The attainable metrological gain depends not only on the reduced spin noise $\xi_{-}^2$, but also on the signal $\aver{|\vec{S}|}$ \cite{Wineland1994} which determines the contrast $C$ of an interferometric measurement. The dominant loss of contrast is due to the scattering of photons into free space during squeezing, which projects atoms into $\spinup$ or $\spindown$. The measured Ramsey contrast as a function of $Q$ is shown in Fig.~\ref{fig:Allan}, together with the a-priori prediction $C{=}C_0\,\textrm{exp}\left(- (Q{/}\tilde{Q}+Q^2)/N \right)$  with the initial contrast $C_0{=}0.97$ in the absence of squeezing as the only fitting parameter. Here, $\tilde{Q}=0.050(3)$ is the independently measured shearing strength per scattered photon. The term $Q{/}\tilde{Q}$ arises from photon scattering into free space, while the second, smaller term accounts for the SSS wrapping around the Bloch sphere.

The metrological gain of a squeezed state is then given by the Wineland parameter $\xi_W^2{=}\xi_{-}^2{/}C^2$ \cite{Wineland1994}. Fig.~\ref{fig:Allan}(a) shows $\xi_W^2$, the measured spin noise reduction $\xi_{-}^2$, and the inferred squeezing $\xi_{st}^2{=}\xi_{-}^2{-}\sigma_d^2$ of the state after subtraction of the measurement resolution $\sigma_d^2$. For $Q \gtrsim 6$ the measured squeezing $\xi_{-}^2$ saturates at $\sigma_d^2$; $Q{=}6.3$ also optimizes the Wineland parameter at $\xi_W^2{=}-6.5(4) \un{dB}$. The inferred squeezing $\xi_{st}^2$ is consistent with the prediction from the model with no free parameters (solid red line), which is limited by the Bloch sphere curvature to $\xi_{st}^2=-15.9(6) \un{dB}$. To our knowledge, this is the first time that the limitation of spin squeezing due to the curvature of the Bloch sphere \cite{Kitagawa1993} has been observed. The inferred metrological gain without readout noise is $\xi_W^2{=}-12.9(6) \un{dB}$.\\
Finally, we directly demonstrate an interferometric measurement with a precision beyond the SQL by implementing a Ramsey sequence with a squeezed, nearly uncertainty-limited input state ($Q{=}3.8$, area $A{=}\sqrt{1{+}F}{=}1.3$). The state is chosen to provide nearly optimum precision gain in the interferometer and in a future OLC \cite{Braverman2018}, see Fig. \ref{fig:Allan}(a). We rotate the squeezed state by $\alpha{=}\pi{/}2-\alpha_\mathrm{-}$ to align the minimal uncertainty along the phase axis, allow the state evolve for a Ramsey time $\tau_R{=}1.5\un{ms}$, and apply a final $\pi{/}2$ rotation to map the accumulated phase onto $S_z$. In Fig.~\ref{fig:Allan}(c) we compare the phase Allan deviation of the SSS (red squares) with that of the CSS (black circles). The Allan deviation for both CSS and SSS Ramsey sequences is derived from 90 sequential measurements. The precision of the CSS interferometer is accurately described by the SQL (black dashed line). The SSS reaches a given precision 4 times faster than a system at the SQL. The main limitation to longer integration times is magnetic field noise (see SM~\cite{SM}). \\
In conclusion, we have demonstrated near-unitary cavity feedback squeezing. Our measurements agree with a model without free parameters that predicts both the area and shape of the squeezed state. The results presented here can be further improved upon in several ways: state detection with a larger applied magnetic field will reduce the Raman scattering and improve the measured spin noise and metrological gain. Alternatively, one can use a squeezing-unsqueezing method \cite{Davis2016,Hosten2016a} that is not limited by the detection quality, and that is also insensitive to the curvature of the Bloch sphere. The intrinsic squeezing of $\xi_{st}^2{=}{-}16 \un{dB}$ is already more than halfway (on a logarithmic scale) between the SQL and the ultimate Heisenberg limit at $\xi_{H}^2{=}{-}30 \un{dB}$ for $N{=}10^3$ atoms. The squeezing performance depends only on the collective cooperativity $N \eta$, and by placing the ensemble at a location in the cavity with higher single-atom cooperativity at constant $N \eta$, i.e., for smaller atom number, the demonstrated performance could already be quite close to the Heisenberg limit. Furthermore, the absolute squeezing will improve with increased collective cooperativity in proportion to $N\eta$. We expect that the spin squeezing can be transferred from the nuclear spin directly to the $\ket{^1S_0} \rightarrow \ket{^3P_0}$ clock transition through an optical $\pi$ pulse, thus enabling optical-clock operation beyond the SQL. Finally, unitary squeezing can also be used to enable quantum information processing with Gaussian states \cite{Weedbrook2012,Opatrny2017}. 
\begin{acknowledgments}
We would like to thank Monika Schleier-Smith, James Thompson, and Mikhail Lukin for valuable discussions.
This work was supported by NSF, DARPA, ONR, and the NSF Center for Ultracold Atoms (CUA). S.~C. acknowledges support as a SNSF Early Postdoc.Mobility fellow. B.~B. acknowledges the support of the Banting Postdoctoral Fellowship.
\end{acknowledgments}

\bibliographystyle{apsrev4-1}
\bibliography{UnitarySqueezing}

\clearpage
\setcounter{figure}{0}\renewcommand\thefigure{S\arabic{figure}}

\setcounter{equation}{0}\renewcommand\theequation{S\arabic{equation}}
\section*{Supplemental Material}
%

\subsection{Technical and experimental details}
Cold \Yb~atoms are prepared in a two-color magneto-optical trap (MOT) \cite{Kawasaki2015} and then cooled further in a single-color MOT on the triplet transition $\term{1}{S}{0} {\rightarrow}\phantom{} \term{3}{P}{1}$ with wavelength $\lambda{=}556 \un{nm}$ and linewidth $\Gamma{/}(2\pi){=}184 \un{kHz}$. The atoms are transported into an asymmetric high-finesse optical cavity \cite{Kawasaki2019} by adjusting the magnetic field of the MOT, and loaded into a one-dimensional optical lattice with wavelength $\lambda_t{=}759 \un{nm}$ and trap depth  $U_0{=} k_B{\times}120 \un{\mu K}$. In order to remove hot atoms with weaker coupling to the cavity mode, the trap depth is lowered to $U_0{/}3$ and restored to $U_0$ over $85 \un{ms}$. The temperature of the remaining $N_\mathrm{tot}{\approx}1500$ atoms is $T{=}(20{\pm}5)\un{\mu K}$.

The asymmetric cavity consists of a large spherical mirror with radius of curvature $R_1{=}25 \un{mm}$, and a slightly elliptical micromirror with an average radius $R_2{=}344\un{\mu m}$ \cite{Kawasaki2019}. The cavity finesse is $\mathcal{F}{=}1.2\times 10^4$ at the probe wavelength $\lambda$, corresponding to a cavity linewidth $\kappa/(2\pi){=}520 \un{kHz}$.
The single-atom cooperativity at an antinode is given by $\eta_0{=}24\mathcal{F}{/}(\pi k^{2}w^{2})$ \cite{Tanji-Suzuki2011}.
At a distance of $0.42 \un{mm}$ from the micromirror, the probe mode waist is $w = 15.1\un{\mu m}$, giving $\eta_0{=}2.4$, which means that the system is in the strong-coupling regime \cite{RJThompson1989kimble,Muenster1999rempe,schleiersmith2019} (see also Ref. \cite{Kawasaki2019} for details). Since $\lambda \neq \lambda_t$, the atoms are inhomogeneously coupled to the probe. As in Refs. \cite{Schleier-Smith2010a,Hu2015}, we define an effective atom number $N{=}N_{tot}\aver{\eta}^2/\aver{\eta^2}{=} \frac{2}{3}N_{tot}$ and effective single-atom cooperativity $\eta{=}\aver{\eta^2}/\aver{\eta}{=}\frac{3}{4} \eta_0$, so that the spin projection noise, measured via the cavity, satisfies the usual relation $\var{N}{=}N/4$ for a CSS. The experiments described below are performed with $N{\approx}1000$, $\eta{=}1.8(1)$, and a collective cooperativity $N\eta{\approx}1800$. The effective cooperativity $\eta$ is confirmed in an independent measurement (see \ref{sec:eta}).

We perform squeezing between the nuclear sublevels $\spinup\equiv \ket{m_I{=}\frac{1}{2}}$ and $\spindown \equiv \ket{m_I{=}{-}\frac{1}{2}}$ of the electronic $\term{1}{S}{0}$ ground state of \Yb. The collective spin state can be represented on a Bloch sphere with radius $S{=}N{/}2$~\cite{Hu2015}. The cavity frequency $\omega_c$ is tuned to be nearly resonant ($\omega_c{-}\omega_a{=} 2\pi \times {-}340 \un{kHz}$) with the $\spinup \rightarrow \ket{\term{3}{P}{1}, m_F=\frac{3}{2}}$ atomic transition with frequency $\omega_a$ in the presence of a magnetic field $B_z{=}13.6\un{G}$ along the cavity axis. $\Nup$ atoms in the state $\spinup$ induce a vacuum Rabi splitting $2g{=}\sqrt{N_\uparrow \eta \kappa \Gamma}$ of the cavity resonance (Fig.~\ref{fig:chirpNgammaSketch}). Near the equator of the Bloch sphere, where $ \Ndown \approx \Nup$, there is also a small dispersive effect from the $\Ndown$ atoms in the state  $\spindown$, suppressed by the Zeeman splitting $\Delta_{z}{=}2\pi\times18.5$~MHz between magnetic sublevels in the excited $\term{3}{P}{1}$ state, with $\Delta_{z}{\gg}{\Gamma,\kappa}$.
To accurately analyze the experiments described below, we need to consider both the near-resonant transition $\spinup \rightarrow \ket{\term{3}{P}{1}, m_F = \frac{3}{2}}$ and the detuned transition $\spindown \rightarrow \ket{\term{3}{P}{1}, m_F = \frac{1}{2}}$.

$S_z$ is determined by detecting $\Nup$ via a measurement of the Rabi splitting $2g$, swapping the populations of $\spinup$ and $\spindown$ with a radiofrequency $\pi$ pulse, and remeasuring the Rabi splitting to give $\Ndown$. From $\Nup$ and $\Ndown$, we determine $S_z=(\Nup-\Ndown)/2$, and $S=(\Nup+\Ndown)/2$ using the two-transition atomic model and the separately measured cavity parameters (\ref{sec:eta}).
The primary quantity of interest, denoted by $\sigma^2 \equiv 2\var{S_z}/S$, is the spin variance $\var{S_z}$ normalized to the CSS noise $\var{S_z}_{CSS}=S/2$. The SQL corresponds to $\sigma^2{=}1$.
Our experimental cavity is frequency-stabilized to the trap laser $\lambda_{t}{=}759 \un{nm}$, whose frequency is stabilized to a stable external reference cavity. The bridging frequency between this cavity and the experimental cavity is set through a sideband generated by an electro-optic modulator (EOM). 
We use the Pound-Drever-Hall technique to lock the cavity to the trap light. 
In order to actively suppress slow drifts of the experimental cavity resonance frequency with respect to the atomic transition we scan the probe light ($\lambda{=}556 \un{nm}$), which is locked to an ultra-stable cavity, and monitor the transmission of the probe light through the cavity. With this we can precisely determine the resonance frequency of the experimental cavity relative to the ultra-stable cavity. We keep this frequency stable from run to run by feeding back through an FPGA circuit on the bridging frequency with an error of less than 15~kHz.

\subsection{Measurement of the atomic state}
\label{sec:AtomicStateMeasurement}
The collective atomic state projection $S_z$ is obtained from the difference $S_z{=}(N_\uparrow-N_\downarrow){/}2$ between the two populations $N_\uparrow$ and $N_\downarrow$ of the states $\spinup=\ket{6s^2~\term{1}{S}{0},~m_I=\frac{1}{2}}$ and $\spindown=\ket{6s^2~\term{1}{S}{0} ,~m_I=-\frac{1}{2}}$.
We first measure the population $N_\uparrow$ of the $\spinup$ state through the vacuum Rabi splitting of the cavity mode $2g{\approx}\sqrt{N_\uparrow\eta\kappa\Gamma}$ occurring when the empty cavity mode frequency $\omega_c$ is resonant with the atomic transition $\spinup \rightarrow \ket{\term{3}{P}{1}, m_F = \frac{3}{2}}$ with frequency $\omega_a$. After that, we apply a radiofrequency (RF) $\pi$-pulse that switches the populations of $\spinup$ and $\spindown$, and remeasure the Rabi splitting which is now proportional to $\sqrt{N_\downarrow}$. We implement the following state measurement sequence ($N_\uparrow^{(1a)}${--}$N_\downarrow^{(1a)}${--}$N_\downarrow^{(1b)}${--}$N_\uparrow^{(1b)}${--}$N_\uparrow^{(2a)}${--}$N_\downarrow^{(2a)}${--}$N_\downarrow^{(2b)}${--}$N_\uparrow^{(2b)}$). The inferred $S_z$ values for the two measurements ($i=1,2$) are $S_z^{(i)}{=}(N_\uparrow^{(ia)}+N_\uparrow^{(ib)}-N_\downarrow^{(ia)}-N_\downarrow^{(ib)}){/}4$. We define the quality of a single measurement as the normalized variance $\sigma_d^2 \equiv \textrm{var}\left( S_z^{(2)}-S_z^{(1)} \right)/S$.
In principle, to determine the state $S_z$ it is sufficient to measure $N_\uparrow$, apply a $\pi{-}$pulse, and measure $N_\downarrow$.
However, in this way the two populations are not measured simultaneously, resulting in additional noise due to atom number decay.
By measuring $N_\downarrow$ and $N_\uparrow$ twice, as described in the previous paragraph, we can eliminate the noise due to atom decay to first order.
\begin{figure}[ht]
\setlength{\unitlength}{1\columnwidth}
    \centering
    \includegraphics[width=.95\columnwidth]{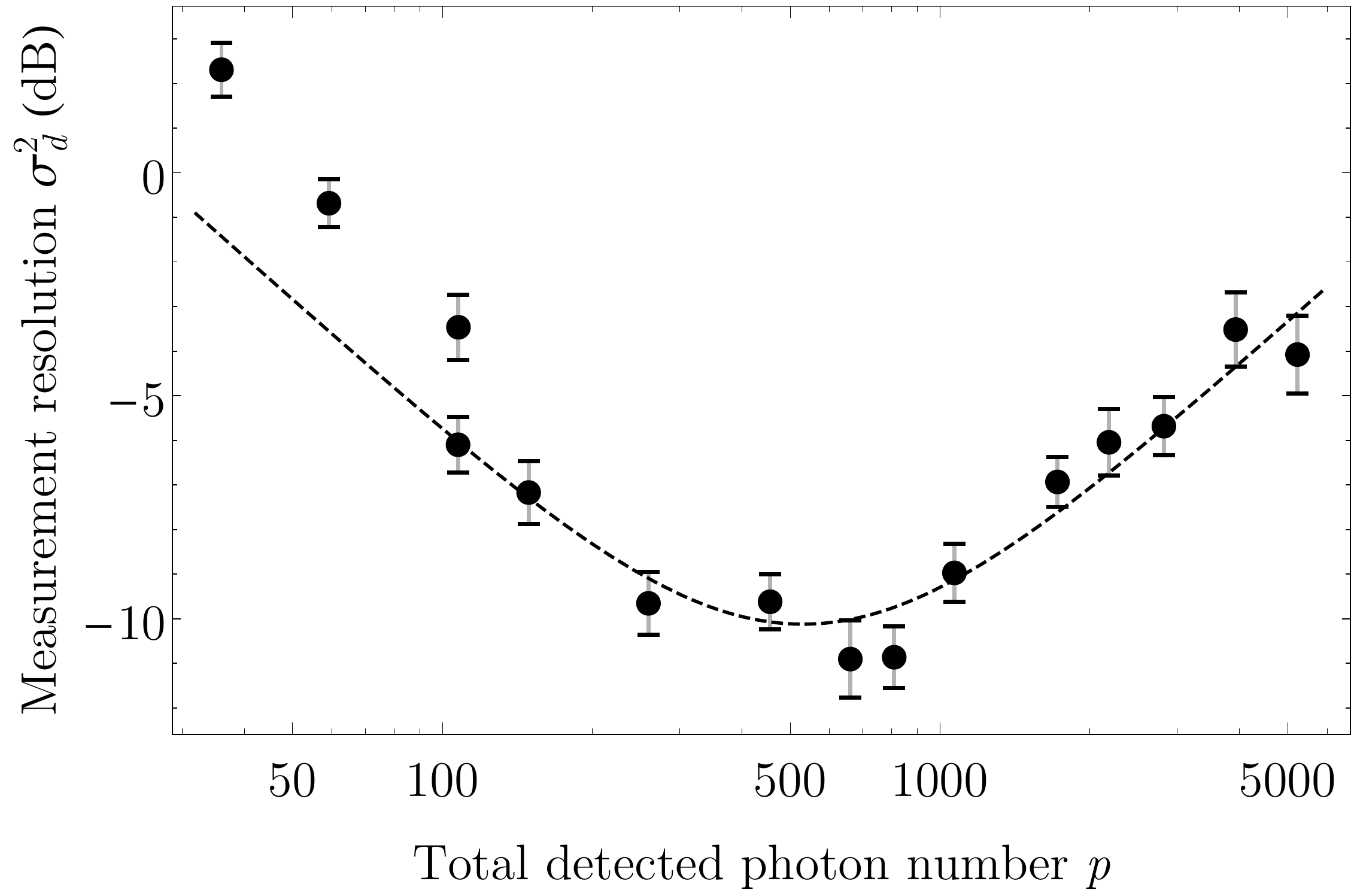}
      \caption{Variance of the difference between two repeated measurements on the same spin state, normalized to the SQL, vs. total detected photon number $p$ per measurement. For small photon number, the measurement quality increases proportionately to $p$. However, at large photon number the variance increases again due to Raman scattering. The dashed line is a fit of form $\sigma^2_d = a/p+bp$ with $a{=}26(9)$ and $b{=}9.2(8){\times}10^{-5}$. The parameters for these measurements are $N{\sim}1000$, $\eta{=}2.0(3)$. 
    }
\label{fig:chirpNgamma}
\end{figure}
Note that because of the existence of the second (detuned) transition $\spindown \rightarrow \ket{\term{3}{P}{1}, m_F = \frac{1}{2}}$, the vacuum Rabi peaks are not exactly symmetric, and we detune the cavity by a small amount $\omega_c-\omega_a=-2\pi\times 340$~kHz from the atomic frequency to cancel the asymmetry to lowest order. We use the exact relation $2g(N_\uparrow,N_\downarrow)$ when determining the populations from the measured Rabi splittings.
\begin{figure}[ht]
\setlength{\unitlength}{1\columnwidth}
    \centering
    \includegraphics[width=.95\columnwidth]{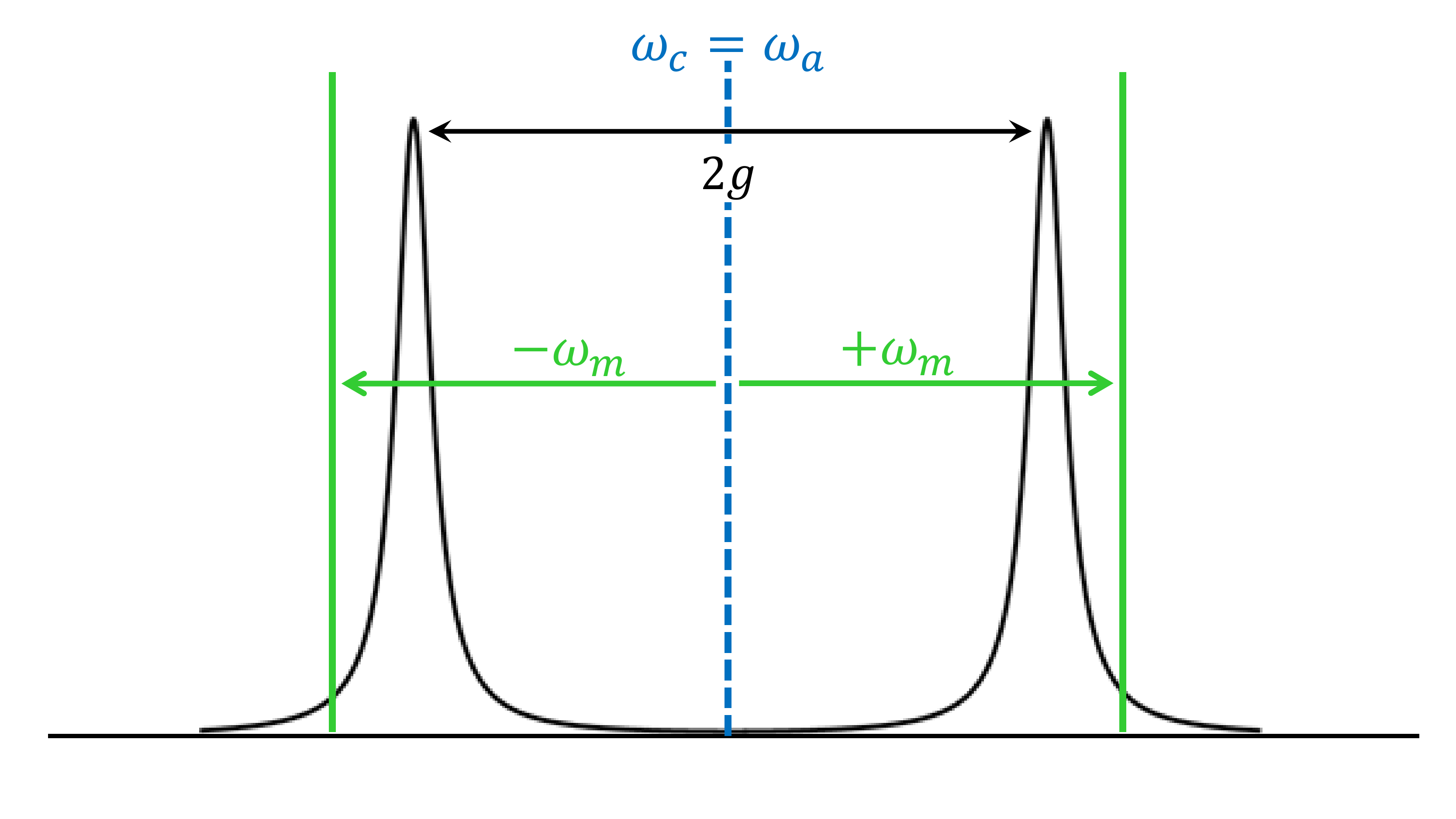}
      \caption{Sketch of the implemented heterodyne measurement. Two laser sidebands (green solid vertical lines) are simultaneously chirped from the frequency of the empty cavity mode (blue dashed vertical light) through the blue and red detuned Rabi peaks, respectively. We use the beating between these two sidebands (beat note at 2$\omega_m$) to measure the Rabi splitting $2g$.
    }
\label{fig:chirpNgammaSketch}
\end{figure}
In Fig. \ref{fig:chirpNgammaSketch} it is shown the Rabi splitting measurement performed by simultaneously sending two laser sidebands, at $\omega_c\pm\omega_m$, into the atom-cavity system. In this way, we can detect both vacuum Rabi peaks simultaneously, while being robust against laser and cavity noise. We perform a linear chirp of the modulation frequency $\omega_m$ from $0$ to $7 \un{MHz}$ in $10 \un{ms}$, and detect the transmitted light that contains a beat note at $2\omega_m$. Information about the atomic state is contained in both the intensity and the phase of the signal at $2\omega_m$. We record the arrival times of the transmitted photons and fit both the intensity and phase; using the phase information in addition to the intensity improves the detection by up to a factor of 4.
As shown in Fig. \ref{fig:chirpNgamma}, detection of more photons improves the $S_z$ detection until Raman scattering increases the $S_z$ noise \cite{Schleier-Smith2010,Hosten2016,Cox2016a,Saffman2009,Chen2014}. In our system, Raman scattering is suppressed to 5\% of Rayleigh scattering due to the small atomic linewidth $\Gamma=2\pi\times184\un{kHz}$ and Rabi splitting $g \approx 2\pi\times 4 \un{MHz}$ compared to the Zeeman splitting $\Delta_z=2\pi\times18.5\un{MHz}$. This enables reasonably good detection at our relatively small atom number $N{\approx}10^3$ compared to $N = 5 \times 10^5$ atoms as used in Refs. \cite{Hosten2016,Cox2016a} which have demonstrated record squeezing. However, our detection is $4 \un{dB}$ worse than the optimum expected for our photon detection efficiency of $15\%$. This is likely caused by the imperfect contrast in the chirp measurement, which hinders the possibility of using the entire phase information acquired in the heterodyne detection.  
\subsection{Measurement of single-atom cooperativity $\eta$ and determination of standard quantum limit (SQL)} 
\label{sec:eta}
The single-atom cooperativity $\eta$ can be calculated from the cavity parameters and the measured position of the atoms along the cavity mode. It can also be experimentally verified from the spin noise, measured via the cavity, as a function of collective cooperativity $N \eta$ \cite{Schleier-Smith2010}. For a state prepared at the equator of the Bloch sphere, the variance of the difference $S_z \eta{=}(N_\uparrow-N_\downarrow)\eta/2$ at the projection noise limit (standard quantum limit (SQL)) is given by
\begin{equation}
    \mathrm{var}\left(\frac{N_\uparrow\eta-N_\downarrow\eta}{2}\right){=}\left({N_\uparrow{+}N_\downarrow}\right) \frac{{\eta^2}}{4}{=}(N\eta) \frac{\eta}{4},
    \label{eq:projectionNoise}
\end{equation}
where $N=\Nup+\Ndown{=}N_{tot}\aver{\eta}^2/\aver{\eta^2}$ is the effective atom number, and $\eta=\aver{\eta^2}/\aver{\eta}$ the effective cooperativity. For atoms uniformly (or randomly) distributed along the length of the cavity, we have $N=\frac{2}{3} N_{tot}$ and $\eta=\frac{3}{4} \eta_0$ \cite{Hu2015}. Here, $N_{tot}$ is the actual number of atoms coupled to the cavity, while $\eta_0$ is the cooperativity at a cavity antinode, given by
\begin{equation}
    \eta_0 = \frac{24 \mathcal{F}}{\pi k^2 w^2}
    \label{eq:CooperativityDefinition}
\end{equation}
for a Gaussian cavity mode. In \eqref{eq:CooperativityDefinition}, $\mathcal{F}$ is the cavity finesse, $k=2\pi / \lambda$ is the wavevector of light resonant with the atomic transition, and $w$ is the $1/e^2$ intensity radius of the cavity mode at the position of the atoms. The asymmetric structure of our optical cavity makes the mode waist $w$ position-dependent \cite{Kawasaki2019}, and equal to $w = 15.1\un{\mu m}$ at the distance of $0.42 \un{mm}$ from the micromirror where we trap the atoms for all experiments described here.
\begin{figure}[ht]
\setlength{\unitlength}{1\columnwidth}
 \begin{picture}(1,0.6)
    \put(.0,0.0){\includegraphics[width=0.95\columnwidth]{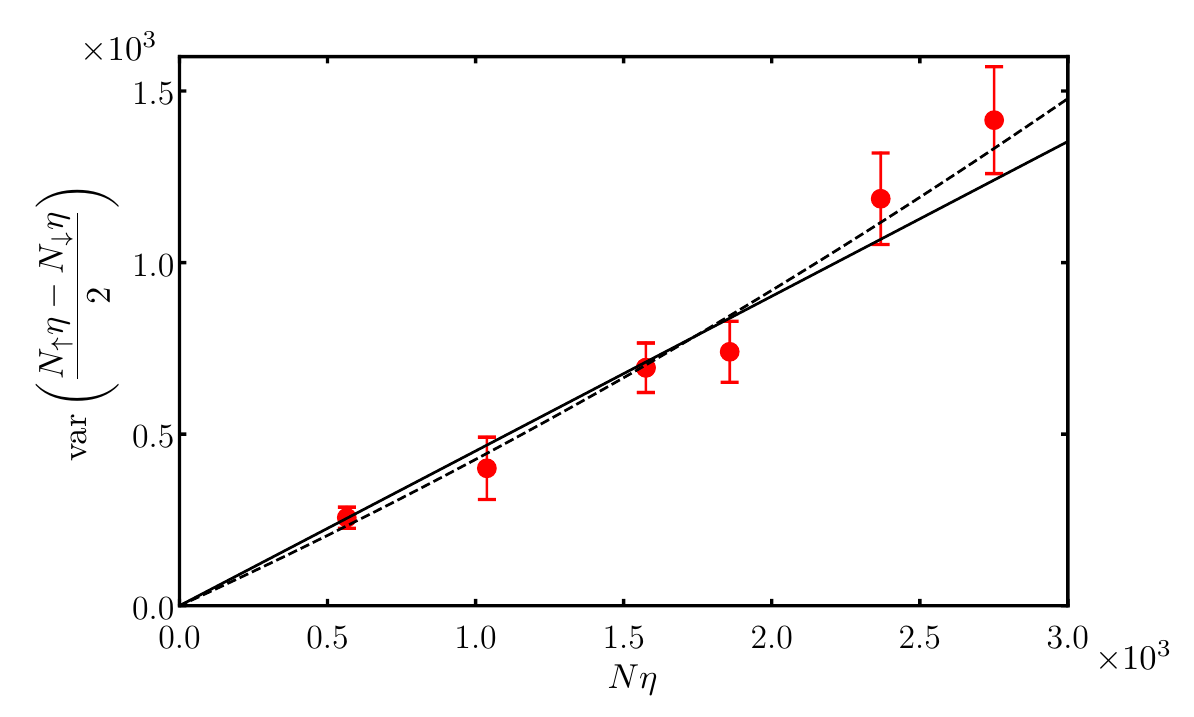}}
    \end{picture}
    \caption{Determination of the effective cooperativity from measured spin noise, see Eq. \eqref{eq:projectionNoise}.
    The solid line represents the data fitted to a linear model, while for the dashed line the model is quadratic.
    }
\label{fig:varianceVSNeta}
\end{figure}
Plotting the measured variance of $\frac{1}{2}(N_\uparrow \eta -N_\downarrow \eta)$ as a function of the measured total cooperativity $N \eta$ yields $\eta/4$ as the slope of the line, as shown in Fig.~\ref{fig:varianceVSNeta}. The result, $\eta{=}1.8(1)$, agrees with the expected value $\eta=1.8(2)$ calculated from the cavity parameters and first principles \cite{Tanji-Suzuki2011}. The SQL in the population difference quadrature is then given by $\var{S_z}{=}S/2{=}N/4$. The linear dependence in Fig. \ref{fig:varianceVSNeta} demonstrates that our system is dominated by quantum noise.
We fitted the data also to a quadratic model (see Fig. \ref{fig:varianceVSNeta}) obtaining $\eta{=}1.6(2)$, which equally agrees with the expected cooperativity.
However, the $p{-}$value of the estimated quadratic coefficient is $0.28$, which means that such a coefficient is consistent with 0.
We thus use the simple linear fit as the best estimator of $\eta$.
\subsection{Generation of RF Pulses}
\label{sec:RFPulseGeneration}
To drive the transition between the two magnetic sublevels $m_I{=}\pm\frac{1}{2}$ of the ground state $\ket{\term{1}{S}{0}, I=\frac{1}{2}}$ of \Yb we use radiofrequency (RF) pulses. The nuclear $g$ factor of \Yb is equal to $g_I = -0.4919(3)$, resulting in a Zeeman shift of $\Delta_{z,I}(m_I) = g_I \mu_N m_I = 2\pi \times - 375 \un{Hz/G} \times m_I$. Thus, the transition frequency $\ket{m_I{=}\frac{1}{2}} \rightarrow \ket{m_I{=}-\frac{1}{2}}$ has a Zeeman shift of $\Delta_{z,I}(\frac{1}{2}) - \Delta_{z,I}(-\frac{1}{2}) {=}2\pi \times -750\un{Hz/G}$. At a typical applied field of $B_z{=}13.6\un{G}$, the resulting splitting (Larmor frequency) is $2\pi \times 10.2 \un{kHz}$. The RF-pluses necessary to drive nuclear spin flips are generated using a single coil which is composed of two independent conductors, generating an oscillating magnetic field in the $\hat{x}$ direction. Each conductor carries the same AC current, but opposite DC currents, to avoid altering the DC magnetic field experienced by the atoms. With an amplitude of $63 \un{A}$ for the alternating current, a Rabi frequency of $208(2) \un{Hz}$ is obtained. More details about the configuration and control of the magnetic field for Rabi pulses can be found in \cite{Braverman2018a}.
In order to be insensitive to the environmental magnetic field variations, we use a CORPSE composite $\pi$-pulse \cite{Cummins2003,Masamitsu2013} in the spin echo sequence. We perform the spin-echo pulse around the $S_x$ axis, i.e., the direction of the average spin vector. The CORPSE pulse consists of the following three pulses \cite{Cummins2003,Masamitsu2013}:
\begin{equation}
\big[\theta_{1}\big]_{[\phi_1]}\big[\theta_{2}\big]_{[\phi_2]}\big[\theta_{3}\big]_{[\phi_3]}
\label{eq:RfPulses sequence}
\end{equation}
In this equation, $\theta_i$ is the pulse area and $\phi_i$ is the relative phase. We choose  $\theta_{1}{=}2\pi + \theta/2 - k$, $\theta_{2}=2\pi - 2k$, and $\theta_{3}{=}\theta/2 - k$, with $k=\mathrm{arcsin}[\mathrm{sin}(\theta/2)/2]$, and $\theta$ is the target rotation pulse area, which in our case is $\theta{=}\pi$. The phases of the three pulses are given by $\phi_{1}=\phi_{2}-\pi{=}\phi_{3}{=}\phi$, where $\phi$ is the phase of the composite pulse. The lengths of the pulses are $5.25 \un{ms}$, $3.75 \un{ms}$, and $0.75 \un{ms}$, respectively. 
The pulse areas are calibrated by Rabi spectroscopy between the nuclear sublevels $m_I=\pm \frac{1}{2}$. The simple $\pi$ Rabi pulse has an efficiency of 98\%. For calibration of the Larmor frequency, we perform Ramsey spectroscopy in the space of the nuclear spin states, with Ramsey times ranging from 1 ms to 50 ms.
\subsection{Experimental sequence for two-color squeezing}
\label{secsequence}
\begin{figure*}[ht]
    \centering
    \includegraphics[width=.8\textwidth]{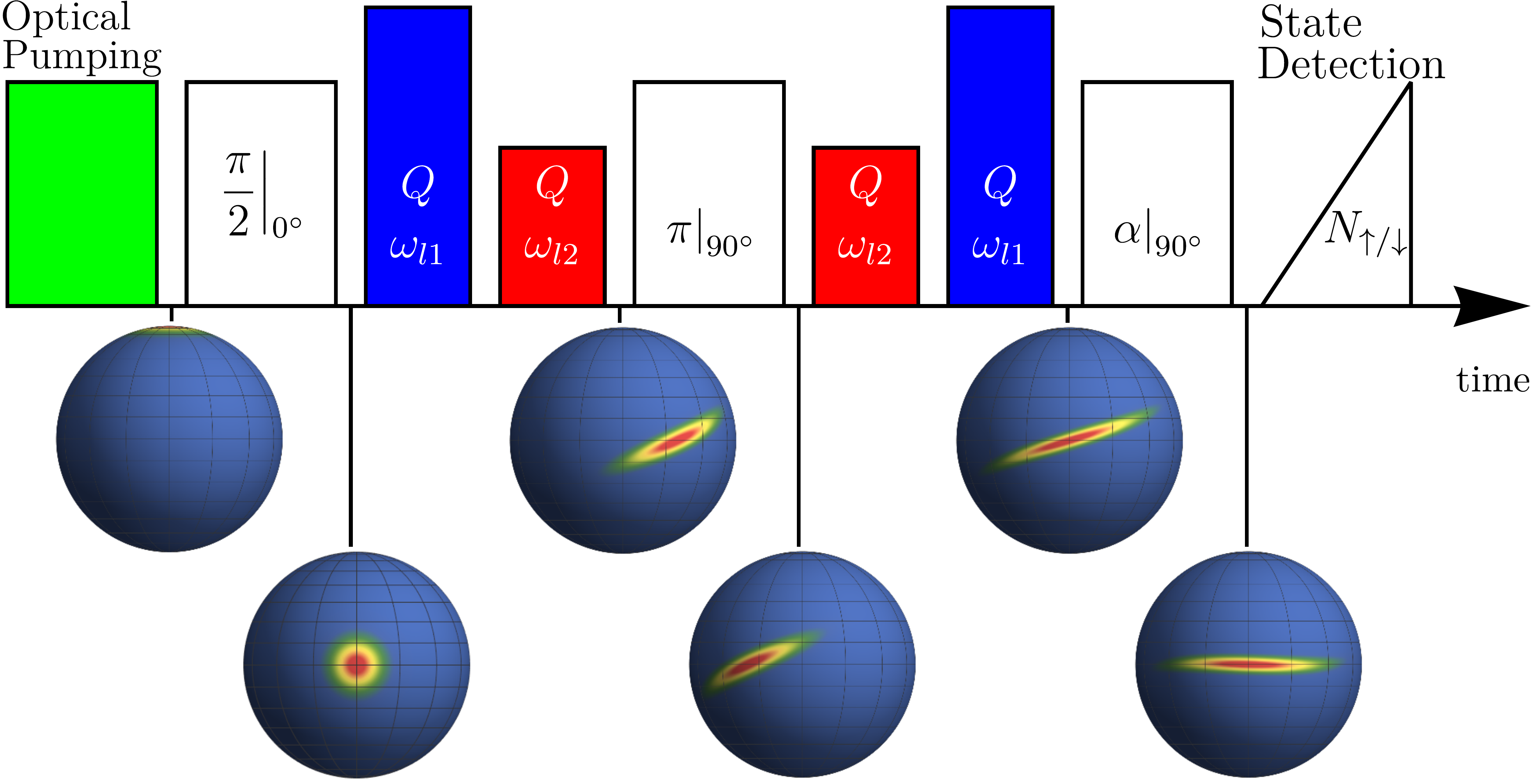}
    \caption{Full experimental sequence for spin squeezing. The subscript in the RF pulses indicates the axis of the rotation. The spheres below the sequence indicate the collective spin state at the corresponding time.}
    \label{fig:fullsequence}
\end{figure*}
Figure~\ref{fig:fullsequence} shows the detailed experimental sequence, a condensed version of which is shown in Fig.~1(d) of the main text. Optical pumping puts all atoms into the $\spinup$ state. A $\pi/2$ pulse (see Sec. \ref{sec:RFPulseGeneration}) prepares a coherent spin state (CSS) along the $+\hat{x}$ direction. Two squeezing pulses are sequentially sent into the cavity, with frequencies $\omega_{l1} = \omega_a + 2\pi{\times}7.33 \un{MHz}$ and $\omega_{l2} = \omega_a - 2\pi{\times}2.00 \un{MHz}$ respectively, and a relative incident power ratio of $P_2{/}P_1{=}0.53$. The shearing induced by a photon at $\omega_{l2}$ is greater than for a photon at $\omega_{l1}$ due to a smaller detuning from atomic resonance. 
At this point, a SSS is already produced, but an $N$-dependent first-order phase shift has displaced the state away from pointing in the $+\hat{x}$ direction. To compensate for this shift, we perform a spin echo sequence, by applying a CORPSE $\pi$ pulse around the $+\hat{y}$ direction, and sending two more squeezing pulses through the cavity, now in reverse order: $\omega_{l2}$ followed by $\omega_{l1}$, with the same power ratio as before. The spin echo sequence also cancels spin dephasing due to inhomogeneous magnetic fields. The final step in the sequence is a rotation of the state around the $+\hat{x}$ direction by a variable angle $\alpha$, followed by the state detection sequence described in Section \ref{sec:AtomicStateMeasurement}.
\subsection{Compensation for Fluctuations in the Total Atom Number}
%
\begin{figure}[hb]
\setlength{\unitlength}{1\columnwidth}
 \begin{picture}(1,.6)
    \put(.05,0.0){\includegraphics[width=.95\columnwidth]{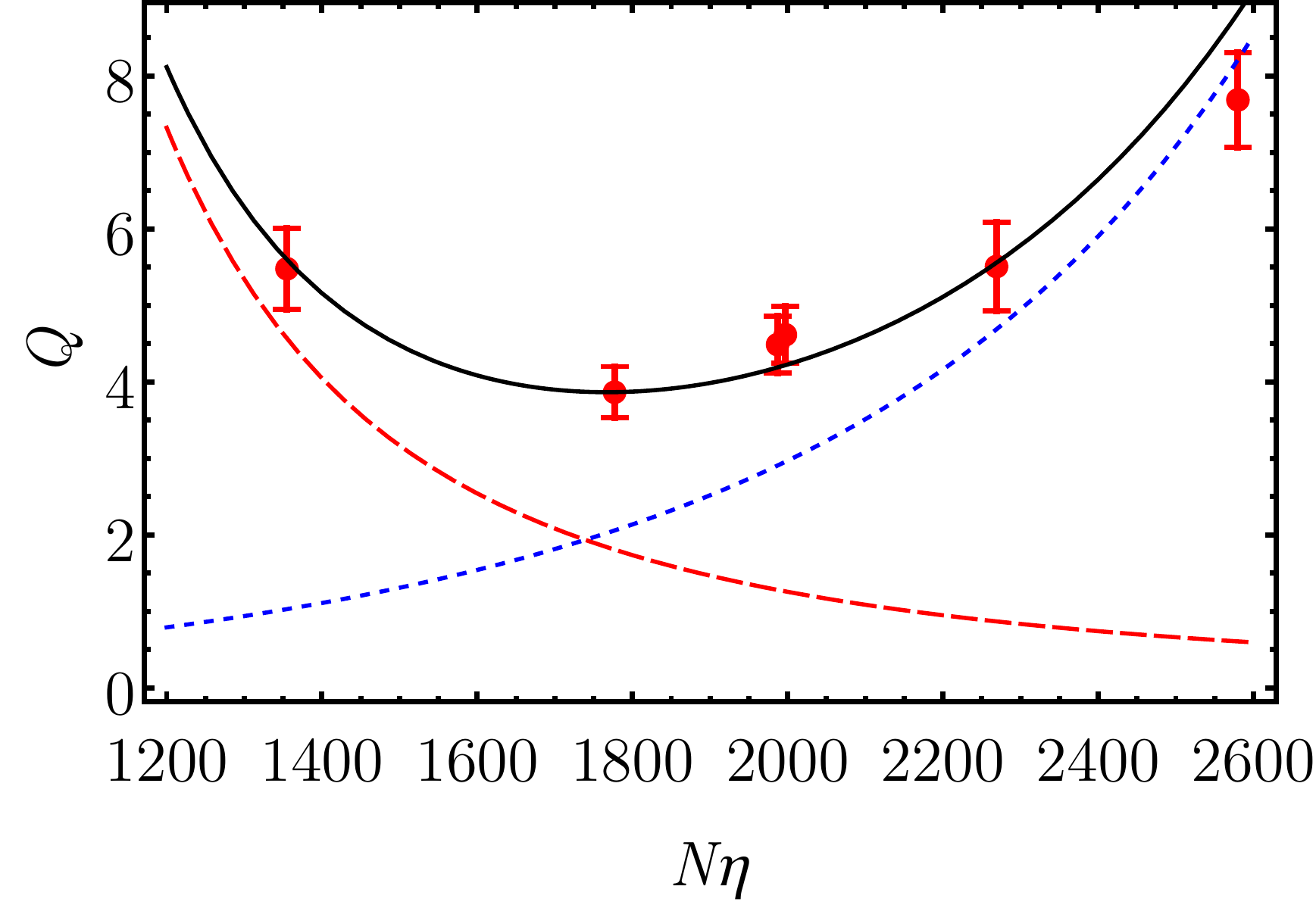}}
    \end{picture}
    \caption{Compensation for total atom number fluctuations measured for $\omega_{l1}=7.334 \un{MHz}$ and $\omega_{l2}=-2.5 \un{MHz}$, $\eta{=}3$. Note that these measurements were taken at slightly different conditions than the other results in the main text.  
    The two squeezing pulses have power ratio $P_2/P_1 = 0.53$. Blue dotted, red dashed and black solid lines stand for theoretical calculation for shearing strength $Q$ due to the $\omega_{l1}$ pulse, the $\omega_{l2}$ pulse and both together, respectively, while the red circles indicate experimental results. Around $N\eta\approx 1800$, $Q$ depends only weakly on the total atom number.
}
\label{fig:QVSNeta}
\end{figure}
For an initial CSS near the Bloch sphere equator (fixed $S_z\approx 0$), fluctuations in the total atom number $N$ change the vacuum Rabi splitting and the shearing strength $Q$ per incident photon. If the atom number $N$ is greater, probing pulses detuned from the atomic frequency by more than the Rabi splitting ($|\omega_{l}-\omega_a| > g \approx 2\pi \times 4 \un{MHz}$) will introduce a larger shearing per incident photon; conversely, pulses with $|\omega_{l}-\omega_a| < g $ will induce less shearing. Thus, by sending two separate probing pulses with appropriately chosen frequencies and relative intensities, it is possible to avoid broadening of the generated squeezed state due to fluctuations in $N$.
To achieve first-order insensitivity to atom number fluctuations around $N\eta=1800$ with $\eta = 1.8$, we choose the parameters ($(\omega_{l1}-\omega_a)/(2\pi){=}7.33 \un{MHz}$, $(\omega_{l2}-\omega_a)/(2\pi){=}-2 \un{MHz}$, and incoming power ratio $P_2/P_1=0.53$. For the conditions in which we obtained the results reported in the main text, the model predicts, a maximum variation of $Q$ equal to $8\%$ in the range $1600 < N\eta < 2000$.
Fig.~\ref{fig:QVSNeta} shows a measurement of the compensation together with a prediction from the theoretical model (see Sec.~\ref{sectheory}) with no free parameters. 
\subsection{Theoretical Model for Squeezing and Measurement Strength}\label{sectheory}
The squeezing is caused by light circulating in the cavity that is influenced by the quantum noise of the atomic spin $S_z$ \cite{Leroux2010,Schleier-Smith2010a}. For $\omega_l, \omega_c, \omega_a$ being the angular frequencies of the incident light, the empty cavity, and the atomic transition, respectively, we define normalized cavity and atomic detuning $x=2(\omega_l-\omega_c)/\kappa$, $y=2(\omega_l-\omega_a)/\Gamma$, and
\begin{eqnarray}
\mathcal{L}_a(y) = \frac{1}{1+y^2} \\
\mathcal{L}_d(y) = -\frac{y}{1+y^2}
\end{eqnarray}
as the absorptive and dispersive Lorentzian lineshapes, respectively. Then in the two-level approximation, where only the state $\spinup$ is coupled to an excited state, the Hamiltonian of the system can be written as \cite{Braverman2019New,Braverman2018a}
\begin{equation}
H = S_z\eta \frac{|\mathcal{E}_c|^2}{\omega_l} \frac{\pi}{\mathcal{F}}\mathcal{L}_d(y)
\end{equation}
Here, $\mathcal{E}_\mathrm{c}$ is the amplitude of the circulating intracavity electric field. This Hamiltonian causes a precession of the spin that is canceled by the spin echo $\pi$ pulse (see Sec.~\ref{secsequence} and Fig.~\ref{fig:fullsequence}). Furthermore, due to the dependence of the cavity field $\mathcal{E}_\mathrm{c}$ on $S_z$, different $S_z$ components experience different phase shifts (one-axis twisting \cite{Kitagawa1993}). The shearing strength $Q$ is defined as the second derivative of the phase shift with respect to $S_z$. Expressing $Q$ by the transmitted photon number $p_\mathrm{tr}$ for a (nearly) one-sided cavity driven through the low-transmission mirror, we obtain for the shearing strength $\hat{Q}$ per transmitted photon the following expression:
\begin{widetext}
\begin{equation}
    \hat{Q}_\mathrm{tr}\equiv\frac{Q}{p_\mathrm{tr}} = - \frac{y}{2\left( 1+y^2\right)^2} \frac{N_\uparrow \eta^2 \left(1+N_\uparrow\eta-xy\right)}{(1+N_\uparrow\eta\mathcal{L}_a(y))^2+(x+N_\uparrow\eta\mathcal{L}_d(y))^2}.
\end{equation}
\end{widetext}
%
Since at fixed parameters the photon number scattered by the atoms into free space is proportional to the transmitted photon number $p_\mathrm{sc}$, we can easily find the shearing strength $\hat{Q}_\mathrm{sc}= Q/p_\mathrm{sc}$ per scattered photon. For our parameters we calculate $\hat{Q}_\mathrm{sc}^\mathrm{(1)}=0.014, \hat{Q}_\mathrm{sc}^\mathrm{(2)} =0.097$ for the blue and red detuned probing pulses, respectively. Taking the input intensity ratio $P_2/P_1=0.53$ and the cavity transmission ratio $T_2/T_1=0.27$ into account, the average shearing strength per scattered photon is $\hat{Q}_\mathrm{sc}=0.024$. 
The excess state broadening $F$ over unitary squeezing arises from the fact that the transmitted and scattered light carries some residual information about the atomic spin $S_z$, and tracing over the light degrees of freedom causes excess antisqueezing of the atomic spin. We define $1+F$ as the factor by which the minimum spin quadrature variance is increased due to excess antisqueezing. It can be shown that $F$ is proportional to the transmitted or scattered photon number. The value $\hat{F}$ per transmitted photon is calculated as \cite{Braverman2019New,Braverman2018a,Kawasaki2017}:
\begin{widetext}
\begin{equation}
    \hat{F}_\mathrm{tr} \equiv\frac{F}{p_\mathrm{tr}} = \frac{2}{\left( 1+y^2\right)^2} \frac{N_\uparrow \eta^2 \left(1+N_\uparrow\eta+ y^2 \right)}{(1+N_\uparrow\eta\mathcal{L}_a(y))^2+(x+N_\uparrow\eta\mathcal{L}_d(y))^2}.
\end{equation}
\end{widetext}
This gives for the excess broadening per scattered photon $\hat{F}_\mathrm{sc}^\mathrm{1}=0.0018$ and $\hat{F}_\mathrm{sc}^\mathrm{2}=0.016$. Taking the intensity ratio into account, we have $\hat{F}_\mathrm{sc}=0.0036$ per scattered photon.
\subsection{Effect of Bloch sphere curvature}
To lowest order in Q, the one-axis twisting Hamiltonian simply induces unitary squeezing with quadrature variances given by Eq. (2). However, due to the curvature of the Bloch sphere, the minimum spin quadrature increases~\cite{Kitagawa1993}, with the next-lowest order term being an increase of the minimum spin quadrature variance by $Q^4/(24 S^2)$ for a homogeneously coupled system~\cite{Leroux2012}. In our inhomogeneously coupled system, this curvature-induced broadening is twice as large: $Q^4{/}(12 S^2)$. This term reproduces the observed broadening at $Q \gtrsim 10$ in Fig.~3 in the main text. The Bloch sphere curvature limits the maximum squeezing to $\xi_-^2 \geq -15.8\un{dB}$.
\subsection{Cavity QED parameters used in this experiment}
The cavity and atomic parameters are summarized in Table~\ref{tab:parameterlist}. For more details regarding the cavity see Ref.~\cite{Kawasaki2019}. Some parameters have changed their values due to aging of the experimental setup and are slightly different from those summarized in Table I of Ref.~\cite{Kawasaki2019}. Note that the cavity linewidth $\kappa$ includes broadening from the relative frequency stability of the cavity and $556 \un{nm}$ probe laser, while the cavity finesse $\mathcal{F}$ is obtained via a ringdown measurement which is sensitive to the properties of the cavity alone. The total detection efficiency $\epsilon$ for a photon initially inside the cavity includes mirror losses, detection path loss, and finite photodetector quantum efficiency.
\newpage

\subsection{Statistical and systematic errors}

\subsubsection{Statistical errors}
The statistical error in the variance estimation with $n$ measurements is given by 
\begin{equation}
    \Delta\sigma^2=\sigma^2\sqrt{\frac{2}{n-1}}.
\end{equation}
Each $\sigma^2_{\alpha}$ in the tomography (see Fig.~2 in the main text) is obtained by collecting more than 100 measurements, resulting in fractional uncertainties smaller than $12\%$.
For each state tomography curve we performed more than 1000 experiments; we used the whole set of data to estimate the state readout $\sigma_d^2$, resulting thus in a fractional uncertainty smaller than $5\%$.
\subsubsection{Systematic errors}
From day to day we observe fluctuations in the estimated state readout $\sigma_d^2$. 
We attribute this to systematic variations of the 556~nm laser frequency on the order of $\pm40$~kHz.
This induces changes in the contrast of the beat note produced in the heterodyne measurement of $S_z$ (see Section \ref{sec:AtomicStateMeasurement}), resulting in fluctuating efficiency of the information derived from the phase of the beatnote at $2\omega_m$ (see Fig.~\ref{fig:chirpNgammaSketch}).
Thus, the $\sigma_d^2{=}9.4(4)$~dB as specified in the main text is dominated by this systematic error.
\subsubsection{Extra considerations on errorbars}
The experiment for each shearing strength Q was performed on a different day.
To infer the spin projection noise and the potential metrological gain, we use the measurement resolution $\sigma_d^2$ obtained on that particular day.
In this case the $\sigma_d^2$ is not affected by daily variations and its uncertainty is only statistical.
Each day the measurement resolution is obtained from ${\geq}1000$ data, resulting in a fractional uncertainty ${\leq} 4.5\%$.
Each $\xi^2_{st}$ and $\xi_W^2$ without readout noise are derived by subtracting the daily value of $\sigma_d^2$ from the measured spin projection noise reduction $\xi_-^2$.
The errorbars of the inferred spin projection noise $\xi^2_{st}$ (see Fig.~4(a) in main text) are obtained by combining this uncertainty with those of $Q$ and $F$ obtained from the tomography measurements (see Fig.~2 in main text).
Finally, the maximal spin projection noise and potential metrological gain are obtained from the minimum of the $\xi^2_{st}$ measured curve.
The resulting values of $\xi^2_{st}{=}15.9(6)$~dB and $\xi_W^2{=}12.9(6)$~dB have uncertainties smaller than the single-measurement uncertainty.

\begin{table}[H]
  \centering
        \begin{tabular}{l|l}
            \toprule
                Atomic wavelength & $\lambda = 555.799\un{nm}$\\
                Atomic linewidth of $\term{3}{P}{1}$ state & $\Gamma/(2\pi)=184(1) \un{kHz}$\\
                Cavity linewidth & $\kappa/(2\pi)=520(15) \un{kHz}$ \\
                Cavity finesse   & $\mathcal{F}=12.2(4) \times 10^3$\\
                Trans. of $R_1{=}25 \un{mm}$ mirror & $30(1) \un{ppm}$ \\
                Trans. of $R_2{=}344 \un{\mu m}$ mirror & $196(5) \un{ppm}$ \\
                Cavity detuning  & ($\omega_c{-}\omega_a){/}(2\pi){=}{-}340(10)\un{kHz}$\\
                Effective cooperativity   &  $\eta=1.8(1)$\\
                Atomic temperature      & $T=20(5) \un{\mu K}$\\
                Lattice depth    & $U_{0}/h=2.5(6)\un{MHz}$\\
                Axial trapping frequency  & $\omega_{ax}/(2\pi)=140(4) \un{kHz}$\\
                Radial trapping frequency  & $\omega_{r}/(2\pi)=1.4(1) \un{kHz}$\\
                Total photon det. eff. & $\epsilon=15(1)\%$\\
            \botrule
        \end{tabular}
    \caption{Summary of most relevant parameters used in this experiment.}
    \label{tab:parameterlist}
\end{table}

\newpage
\end{document}